\documentclass[twoocolumn,floats,floatfix,superscriptaddress,aps,prl]{revtex4-1}
\usepackage{amsfonts}
\usepackage{amssymb}
\usepackage{amsmath}
\usepackage{verbatim}

\usepackage{calc}
\usepackage{graphicx}
\usepackage{bm}
\usepackage{float}

\usepackage[normalem]{ulem}
\usepackage{amsmath,amssymb}
\usepackage{multirow}
\usepackage{xcolor,soul}
\usepackage{graphicx}
\usepackage{hyperref}
\usepackage{epstopdf}

\def\SL{\textcolor{black}}
\def\blue{\textcolor{black}}

\begin{document}

\author{Ching Hua Lee}
\email{phylch@nus.edu.sg}
\affiliation{Department of Physics, National University of Singapore, Singapore 117542}
\affiliation{Institute of High Performance Computing, A*STAR, Singapore 138632}
\author{S. Longhi} 
\email{stefano.longhi@polimi.it}
\affiliation{Dipartimento di Fisica, Politecnico di Milano and Istituto di Fotonica e Nanotecnologie del Consiglio Nazionale delle Ricerche, Piazza L. da Vinci 32, I-20133 Milano, Italy}
\affiliation{IFISC (UIB-CSIC), Instituto de Fisica Interdisciplinary Sistemas Complejos - Palma de Mallorca, E-07122 Spain}

\title{Ultrafast and Anharmonic Rabi Oscillations between Non-Bloch-Bands}
\normalsize
\date{\today}

\bigskip
\vspace{1cm}
\begin{abstract}
\begin{large}
{\bf Abstract}\\
\end{large}

\vspace{0.1cm}
Rabi flopping between Bloch bands induced by a weak ac resonant field is a coherent effect involving interband transitions. Here we consider the fundamental processes of emission/absorption of quanta and Rabi oscillations in non-Hermitian two-band lattices exhibiting unbalanced non-Hermitian skin effect, and unveil an unprecedented scenario of Rabi flopping.  The effective dipole moment of the transition - usually considered a bulk property - is however strongly dependent on boundary conditions, being greatly enhanced with increased Rabi frequency only when open boundaries are present. As the field strength is increased, Rabi oscillations rapidly become anharmonic, and transitions cease to be vertical in the energy-momentum plane until the system enters into an unstable regime (complex quasi-energy spectrum) due to secular amplification channels. Remaining stable even in the presence of complex energies, Rabi oscillations provide a vivid illustration of how the competition between non-Hermitian, non-local and Floquet influences can result in significant enhancements of physically measurable quantities.
\end{abstract}

\maketitle

\vspace{1cm}
\begin{large}
{\bf Introduction}\\
\end{large} 
--------------------------------------------------------------------------------------------------------------------------------------------------------

\vspace{0.5cm}
The coherent dynamics of electrons in crystalline potentials under time-periodic driving fields is at the heart of such major phenomena as photon-assisted transport, Rabi oscillations, dynamic localization, and super-Bloch oscillations \cite{r1,r2,r3,r4,r5,r6,r7,r8,r9,r10,r10b,r11,r12,r13,r14,r15,r16,r17,r18,r19}. Strong ac fields can modify the band structures of materials  and alter the corresponding internal structure
of the electronic wave functions \cite{R1,R2,R3,R4,R5,R6,R7,R8,R9}, whereas weak ac fields in resonance with two bands of the crystal can induce interband transitions, e.g. absorption and emission of quanta from the field.  Momentum conservation ensures
that direct transitions must be vertical in the $(k,E)$ plane during absorption and emission, where $k$ and $E$ are the quasi-momentum and energy of the electron. Under coherent dynamics,
periodic electron flopping, i.e. Rabi oscillations (ROs) between two Bloch bands, can be observed \cite{r4,r5,r8}. The characteristic frequency of the Rabi flopping is proportional to the electric-dipole moment of the transition and the field strength. {In condensed matter systems, dephasing effects generally prevent the observation of Rabi flopping. For such a reason, ROs have been observed mostly in synthetic lattices, such as in cold atoms and photonic crystals \cite{r5,r14}}.\\
Topological properties, transport and phase transitions in non-Hermitian crystals, i.e. described by an effective non-Hermitian Hamiltonian, have attracted a huge interest in the past few years \cite{r20,r21,r22,r23,r24,r25,r24t,r26,r27,r28,r29,r29b,r29c,li2019geometric,r30,rt1,rt2,r30o,r30b,r30c,r30d,r31,lee2019unraveling,yoshida2019non,yoshida2019mirror,yang2019auxiliary,zhang2020non,liu2020non,borgnia2020non,uffa0,uffa1,uffa2,lee2019unraveling,mu2019emergent,Berg,yi2020non}. In such crystals, the energy spectrum is  strongly sensitive to perturbations, and largely differs under periodic (PBC) and open (OBC) boundary conditions. In systems with OBC the bulk states can get squeezed toward the lattice edges (non-Hermitian skin effect \cite{r21,r22,r23,r24,mu2019emergent,r25,lee2019unraveling}), and the bulk-boundary correspondence based on Bloch band topological invariants generally fails to predict topological edge states. To correctly describe energy spectra and topological invariants in crystals with OBC one needs to extend Bloch band theory so as the quasi-momentum becomes complex and varies on a generalized Brillouin zone (GBZ) \cite{r21,r22,r24,r29,lee2019unraveling}. Bloch and non-Bloch bands show different energy spectra and can undergo different symmetry breaking phase transitions. As major attention is currently devoted to study the topological properties and related symmetries in several non-Hermitian models, the impact of the skin effect on bulk transport properties in non-Hermitian lattices driven by external fields remains so far largely unexplored.

In this work, we show how non-Hermitian influences can disclose a scenario fully distinct from common Rabi flopping, hosting novel features such as enhancement of the effective dipole moment arising from the non-Hermitian skin effect, non-vertical transitions and anharmonic ROs, hence providing unprecedented freedom in controlling the frequency and anharmonicity of ROs. While non-Hermitian systems with complex eigenenergies have often been considered to be of limited experimental interest, since complex eigenenergies seem to lead to rapid decay or divergences, ROs can be sustained without gain or loss, even if the eigenspectrum is complex. This thus greatly expands the scope by which Rabi oscillations can be controlled or engineered, achieving Rabi frequencies that are orders of magnitudes higher than allowed by the bare dipole moments. More generally, our study sheds light on how non-Hermiticity further enriches the already vibrant field of Floquet dynamics \cite{zhou2018non,zhou2019dynamical,zhou2020non,lee2020quenched} beyond merely causing gain or attenuation.

\vspace{1cm}
\begin{large}
{\bf Results}\\
\end{large}
--------------------------------------------------------------------------------------------------------------------------------------------------------

\vspace{0.5cm}
\noindent{\bf {Rabi oscillations in a non-Hermitian lattice}} \\
\\
Let us consider a minimal non-Hermitian 1D system comprising two identical chains (sublattices) $H_O$ coupled by an inter-chain coupling term $H_C$. Acting on the system is a weak ac field $F(t)=F_0\cos (\omega t)$ that staggers the energy of the two sublattices and drives the ROs. The system is thus described by a 2-component Hamiltonian in real space
\begin{equation}
H(t)=\left( 
\begin{array}{cc}
H_O & H_C\\
H_C & H_O
\end{array}
\right)-d_yF(t)\left( 
\begin{array}{cc}
\mathcal{I} & 0 \\
0 & \SL{-} \mathcal{I}
\end{array}
\right),
\end{equation}
where $2d_y$ is the spatial separation between the two chains of length $N$, and $\mathcal{I}$ is the $N \times N$ identity matrix [Fig. 1(a)]. 
Diagonalizing \SL{$H(t)$} in the zero field limit via a basis transformation $ H\rightarrow U^{-1}HU$ with $U=\frac1{\sqrt{2}}\left( \begin{array}{cc}
1 & 1 \\
1 & -1 
\end{array}\right) $, our system takes the form of two effective but \emph{inequivalent} chains $H_\pm=H_O \pm H_C$ that are coupled by the oscillatory field $F(t)$ \SL
{[Fig.1(b)]}. In this new basis built from the symmetric and anti-symmetric sectors, a two-component state $|\psi(t)\rangle=(|\mathbf A\rangle,|\mathbf B\rangle)^T$ obeys the dynamical evolution equation $i\frac{d\psi}{dt}=H\psi$ which read
\begin{equation}
i \frac{d}{dt}
\left(
\begin{array}{c}
\mathbf{A} \\
\mathbf{B}
\end{array}
\right)=
\left[\left( 
\begin{array}{cc}
H_+ & 0 \\
0 & H_-
\end{array}
\right) -d_yF(t)
 \left( 
\begin{array}{cc}
0 & \mathcal{I} \\
\mathcal{I} & 0
\end{array}
\right)\right]
\left(
\begin{array}{c}
\mathbf{A} \\
\mathbf{B}
\end{array}
\right).
\label{AB}
\end{equation}
To solve Eq.~(\ref{AB}), we expand $|\mathbf A\rangle$ and $|\mathbf B\rangle$ in terms of the right eigenvectors $|u^{R}_{+,n}\rangle$ and $|u^{R}_{-,n}\rangle$ respectively defined by $H_\pm|u^{R}_{\pm,n}\rangle =E_{\pm,n}|u^{R}_{\pm,n}\rangle$:
\begin{eqnarray}
|\mathbf{A}\rangle & = & \sum_n \alpha_n(t)  e^{-iE_{+,n} t} |u^{R}_{+,n}\rangle\\
|\mathbf{B}\rangle & = & \sum_n \beta_n(t) e^{-iE_{-,n} t} |u^R_{-,n}\rangle.
\end{eqnarray}
Upon substituting into Eq.~(\ref{AB}) and left multiplying by left eigenvectors defined by $H^\dagger_\pm|u^{L}_{\pm,n}\rangle =\SL{E_{\pm,n}^*} |u^{L}_{\pm,n}\rangle$ and obeying the biorthogonal normalization $\langle u^L_{\pm,n}|u^R_{\pm,l}\rangle = \delta_{nl}$, we obtain coupled equations describing the evolution of the amplitude probabilities $\alpha_n(t)$ and $\beta_n(t)$ of the symmetric/antisymmetric sectors:
 \begin{eqnarray}
 i \frac{d \alpha_n(t)}{dt} & = & -d_y F(t)  \sum_l \Gamma_{n,l} \beta_l(t) e^{i(E_{+,n}-E_{-,l})t} \label{alphabeta1}\\
i \frac{d \beta_l(t)}{dt} & = & -d_y F(t)  \sum_n G_{l,n} \alpha_n(t) e^{i(E_{-,l}-E_{+,n})t}, \qquad 
\label{alphabeta2}
\end{eqnarray}
where $\Gamma_{n,l}=\langle u^L_{+,n}|u^R_{-,l}\rangle$ and $G_{l,n}=\langle u^L_{-,l}|u^R_{+,n}\rangle$. Without any restriction on the forms of $H_\pm=H_O \pm H_C$, Eqs.(\ref{alphabeta1}) and (\ref{alphabeta2}) generically describe wildly fluctuating dynamics that is generically aperiodic \SL{with complex quasi-energy spectrum}. 
\blue{To investigate ROs, we specialize to cases where well-defined oscillations exist between two chosen eigenstates $|u^R_{+,n}\rangle$ and $|u^R_{-,l}\rangle$ \SL{having the same growth/decay rate, i.e. vanishing imaginary part of $E_{+,n}-E_{-,l}$ [Fig.1(c)]}, and make the crude rotating wave approximation (RWA), assuming as usual that $F(t)$ is modulated at resonance $\omega=\omega_{nl} \equiv E_{+,n}-E_{-,l}$ and the \SL{ Rabi frequency is much smaller than $\omega$. Neglecting all non-resonant and cross-coupling terms}, from Eqs.(\ref{alphabeta1}) (\ref{alphabeta2}) harmonic oscillator equations for the coupled amplitudes $\alpha_n$ and $\beta_l$ are obtained, namely  $\frac{d^2\alpha_n(t)}{dt^2}+\frac1{4}(d_y^2F_0^2\Gamma_{n,l}G_{l,n})\alpha_n(t)=0$ (and similarly for $\beta_l(t)$). The Rabi frequency is thus
\begin{equation}
\Omega_R=d_yF_0\sqrt{|\Gamma_{n,l}G_{l,n}|}=d_yF_0\sqrt{|\text{Tr}[P^+_nP^-_l
]|}
\label{OmegaR}
\end{equation}
where $P_\mu^\pm=|u^R_{\pm,\mu}\rangle\langle u^L_{\pm,\mu}|$ is the biorthogonal projector onto the $\mu$-th eigenstate of $H_\pm=H_O\pm H_C$ \cite{note1}. Note that ROs can occur even in the \emph{absence} of a real spectrum, as long as $E_{+,n}-E_{-,l}$ is real. Under PBCs, this reality condition simplifies to the requirement that $H_C$ has a real spectrum.}
\blue{ \SL{Equation (7) shows that the Rabi frequency $\Omega_R$ is proportional to the {\em effective} dipole moment $\mu^{(eff)}_{n,l} \equiv d_y\sqrt{\text{Tr}[P^+_nP^-_l]}$, which can be enhanced in a non-Hermitian system}. In the Hermitian case, this is not possible as $\text{Tr}[P^+_nP^-_l]=|\langle u_{+,n}|u_{-,l}\rangle|^2\leq 1$, with overlap integrals bounded above by unity. But in non-Hermitian cases, eigenstates are biorthogonally normalized, and there are two scenarios where $\text{Tr}[P^+_nP^-_l]$ can be very large: (i) near an exceptional point and (ii) in the presence of boundary eigenmode accumulation, also known as the non-Hermitian skin effect. For (i), exceptional points have been known to harbor pronounced sensitivity due to the defective nature of their eigenspaces, and the dipole moment amplification is expected \SL{\cite{r17}}. But more interesting is (ii), where the effective dipole moment and hence Rabi frequency can be controlled  just  by changing boundary conditions.}
\SL{Moreover, when considering ROs in the $(k,E)$ plane, in the non-Hermitian case boundary conditions drastically {challenge} the common wisdom that transitions has to be vertical and harmonic, as discussed below.} 

\vspace{0.5cm}
\noindent{\bf {Boundary-driven ultrafast and non-vertical Rabi Oscillations}} \\
\\
\begin{figure}
  \includegraphics[width=120mm]{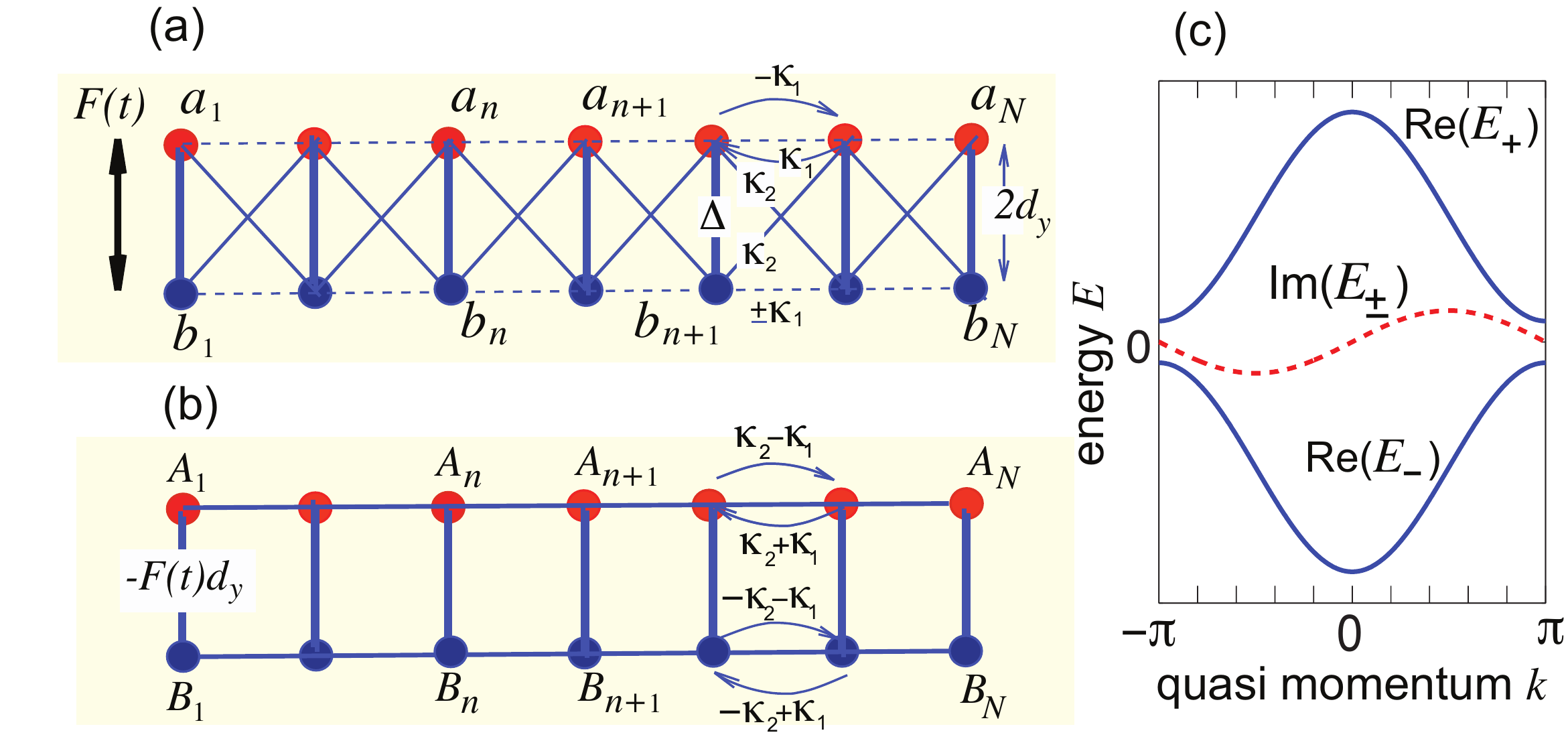}\\
   \caption{(color online) \SL{(a). Schematic of an ac-driven  non-Hermitian lattice. The dashed bonds denote a non-Hermitian hopping, with amplitudes $\pm \kappa_1$ depending on the direction (left/right) of the hopping. (b) Equivalent lattice after the basis transformation. The system is basically equivalent to two Hatano-Nelson chains, shifted in energy by $\pm \Delta$ and side-coupled by the oscillating bond $-d_yF(t)$. The skin modes under OBC are squeezed toward opposite edges in the two chains with the same skin length. (c) Energy bands under PBC in the undriven case ($F=0$).} } 
\end{figure}
\SL{The eigenvectors $|u^{R,L}_{\pm,n}\rangle$, and thus the scalar products $\Gamma_{n,l}, G_{lml}$ defining the interband transitions, depend on the boundary conditions, and differ for PBC and OBC. Under PBC, $\langle x| u^{R,L}_{\pm,n}\rangle=\exp(ik n x)$ are plane waves with quantized quasi-momentum $k_n= 2 n \pi /N$, and thus $\Gamma_{n,l}, G_{n,l}$ vanish for $n \neq l$, i.e. the ac field can induce only vertical transitions in $(k,E)$ plane, both for Hermitian and non-Hermitian systems [Fig.2(a)]. Correspondingly, ROs are always harmonic and there is not any enhancement of the dipole moment ($\mu^{(eff)}_{n,n}=d_y$}).
Under OBC, the quasi-momentum $k$ is not anymore a good quantum number \cite{Viola,Viola1,Viola2,Viola3}, however one can still diagonalize $H_{\pm}$ in real space and the bulk energy spectrum (with the exception of isolated states) can be obtained from the GBZ \cite{r21,r24,r29}, i.e. from the analytic continuation of the Bloch energy bands $E_{\pm}(k)$ where $k$ becomes complex and spans, for each band, a path in complex plane, as detailed in the Methods (see also  \cite{lee2019unraveling}). Here we focus our attention to the most interesting case where $H_{\pm}(k)$, under PBC, can be derived from the {\it same} momentum-space Hermitian Hamiltonian $H_H(k)$ via uniform complex momentum deformations~\cite{Suppl} 
$k\rightarrow k-ih_\pm$, i.e. $H_{\pm}(k)= Q_{\pm}H_H(k-ih_{\pm})$ with $Q_{\pm}$ and $h_{\pm}$ real parameters. This case applies to the typical scenario of ROs in lattices with nearest-neighbor (NN) hopping \cite{r4}, in which the non-Hermitian deformations $h_{\pm}$ are introduced by synthetic imaginary gauge fields \cite{r24,r32,r33,r34}. As explained in the Methods, under OBC $H_{\pm}$ show an entirely real energy spectrum and share the {\em same} eigenmodes with quantized wave number $k_{+,n}=k_{-n} \equiv k_n={\frac{n\pi}{N+1}}$, {$n=1,2,...,N$}. In the Hermitian limit $h_1=h_2=0$, one has $\Gamma_{n,l}=G_{n,l}=0$ for $l \neq n$ and ROs occur under resonance driving between eigenmodes in the two bands with the same quasi-momentum $k_n=k_l$ [Fig.2(a)]. In other words, ROs in the Hermitian limit are the same for PBC and OBC, and transitions remain vertical in the $(k,E)$ plane. This is not the case for non-Hermitian lattices, where boundary conditions affect the dynamics tremendously  as non-vertical transitions are allowed [Fig.2(b)]. 

To illustrate this point, let us consider the NN Hamiltonian $H_{{H}}(k)=2 \sqrt{\kappa_2^2-\kappa_1^2} \cos k+ \Delta$, with $\Delta$ and $\kappa_2 > \kappa_1$ real and positive parameters, and let us assume $Q_{\pm}= \pm 1$ and $h_{\pm}= \mp h$ with $h=(1/2) \log[(\kappa_2+\kappa_1)/(\kappa_2-\kappa_1)]$. The resulting Hamiltonians $H_{\pm}$ read
\begin{equation}
H_{\pm}(k)=2 i \kappa_1 \sin k \pm ( \Delta+2 \kappa_2 \cos k)
\end{equation}
corresponding to {the physical single chain Hamiltonian} $H_O=2 i \kappa_1 \sin k$ and {inter-chain coupling}  $H_C=\Delta + 2 \kappa_2 \cos k$; see Fig.~1(a). After the basis transformation,
the two decoupled lattices $H_{\pm}$ are two Hatano-Nelson chains \cite{r32,r33,r34} with asymmetric left/right hopping and shifted in energy by $\pm \Delta$; see Fig.~1(b). The skin effect squeezes the bulk modes for the two bands towards opposite boundaries. To have well-spaced bands for ROs, we assume that the two bands are separated by a wide gap, i.e. we assume that  $\Delta\gg\kappa_2$. 
 \begin{figure}[htbp]
  \includegraphics[width=100mm]{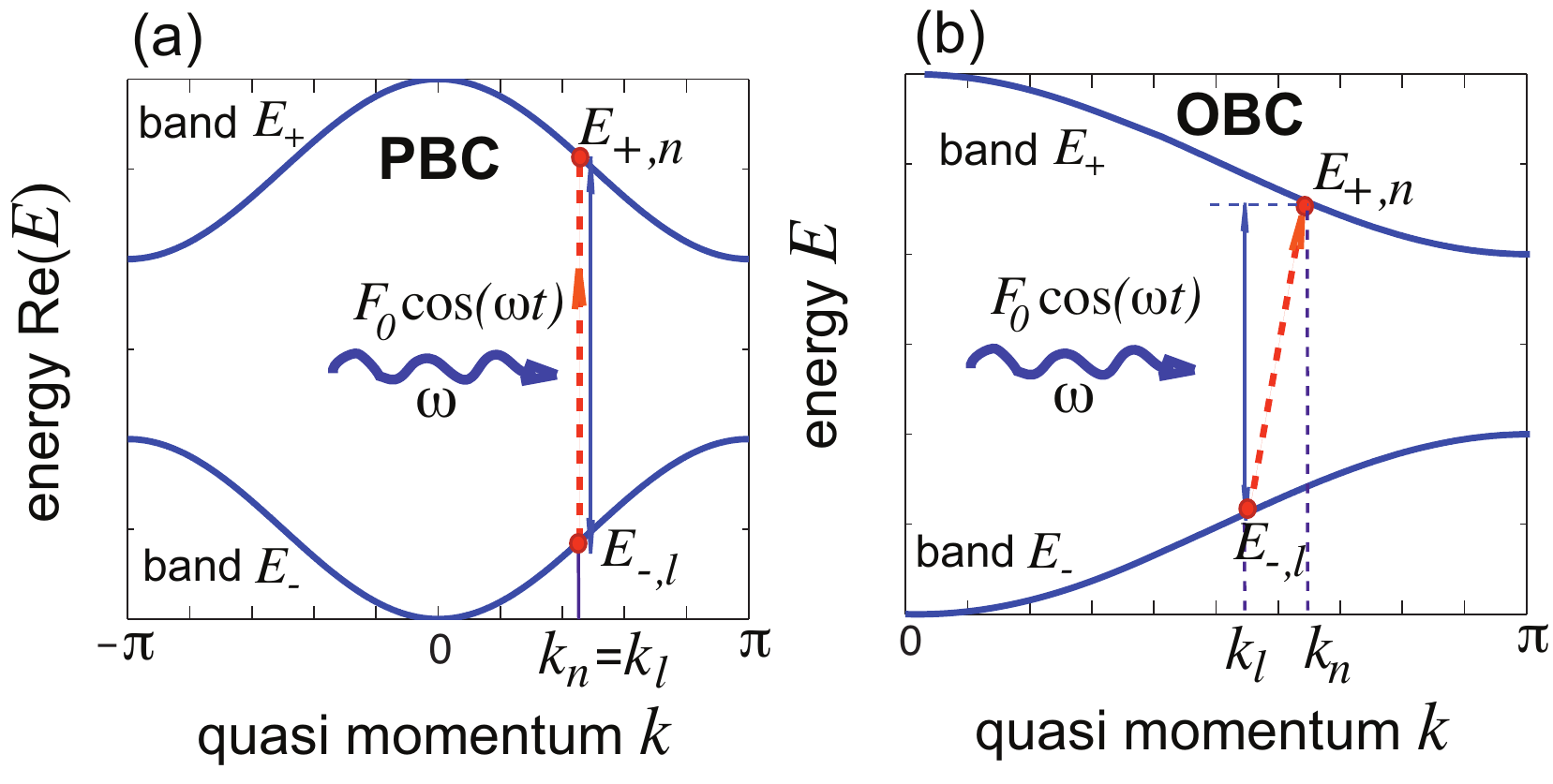}\\
   \caption{(color online) \SL{(a) Field-induced transitions between Bloch bands in a system with PBC. The allowed transitions are vertical in the $(k,E)$ plane. The resulting ROs are always harmonic. (b) Under OBC, transitions between non-Bloch-bands can be oblique in the $(k,E)$ plane. Mixed transitions can lead to anharmonic ROs.}} 
\end{figure}
\begin{figure}[htbp]
  \includegraphics[width=120mm]{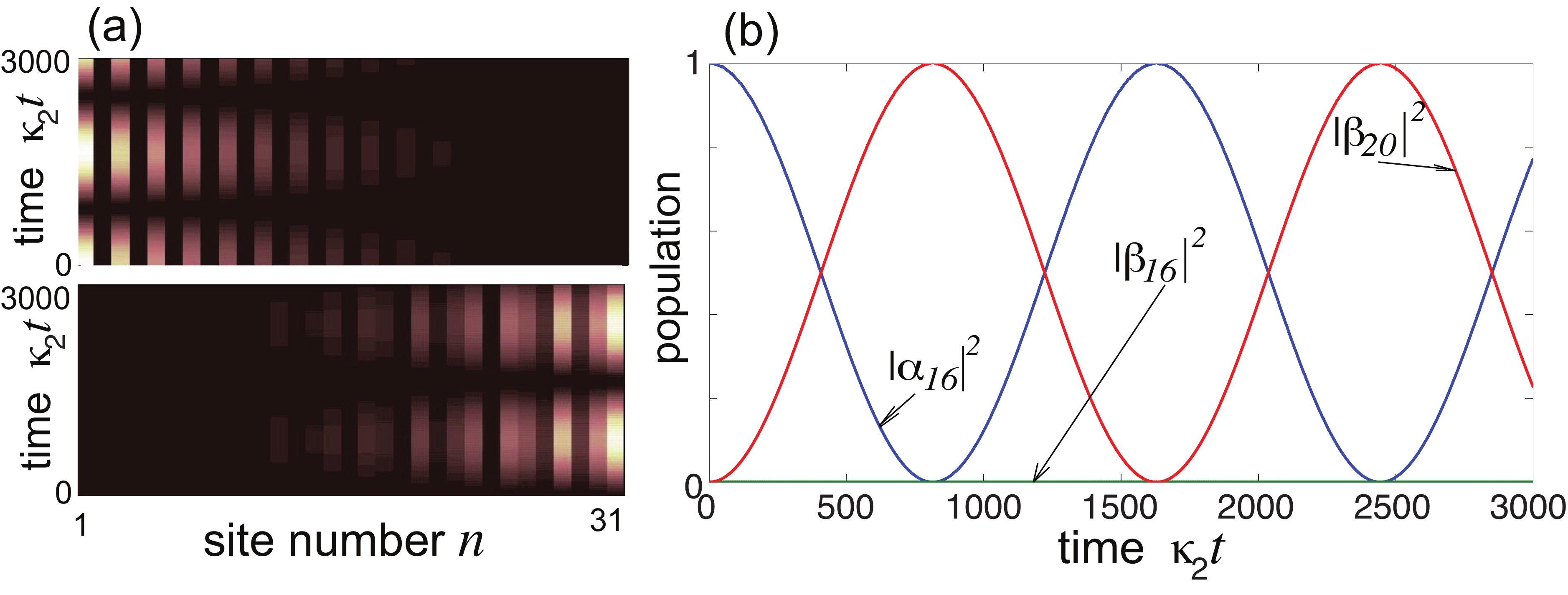}\\
   \caption{(color online)  \SL{Non-vertical ROs in the lattice of Fig.1 comprising $N=31$ unit cells with OBC for $\Delta / \kappa_2=3$, $F_0 / \kappa_2 =0.005$ and $\kappa_1 / \kappa_2=0.1$. The initial state is the bulk eigenstate $ | \mathbf{u}^{R}_{+,n} \rangle$ with quasi-momentum $k_n= n \pi /(N+1)$ and $n=(N+1)/2=16$. The modulation frequency $\omega$ is tuned to put in resonance the modes $k_n$ and $k_l$ (with $l=n+4=20$) in the upper and lower bands. (a) Evolution of the real-space occupation probabilities $|A_n|^2$ and $|B_n|^2$ (upper and lower panels). Different $k$ modes are visible from different fringes in upper and lower panels. (b) Evolution of the occupation probabilities $|\alpha_{16}|^2$ and $|\beta_{20}|^2$. Also the probability $|\beta_{16}|^2$ is shown, which remains almost zero because of chosen ac modulation frequency.}}
\end{figure}
\blue{Under PBCs, ROs \SL{occur only for vertical transitions $k_n=k_l$ [Fig.2(a)], with corresponding eigenmodes displaying the same lifetime [Fig.1(c)]. But under OBC the spectra of the two non-Bloch bands are entirely real and non-vertical transitions are allowed. The OBC energy spectra are readily obtained from the one of $H_H$ and read}
\SL{
\begin{equation}
E_{\pm,n} = \pm \left[ 2 \sqrt{\kappa_2^2-\kappa_1^2} \,\cos k_n + \Delta \right]  
\end{equation}
with $k_n= n \pi /(N+1)$ and $n=1,...,N$. To highlight the appearance of non-vertical transitions,} let us compute the biorthogonal eigenbasis of $H_{\pm}$ under OBC, which are identical upon left/right interchange. We have
\begin{equation}
\langle x|u^R_{\pm ,n}\rangle  =  e^{\pm h(N+1)/2}\sqrt{ \frac{2}{N+1}} \sin \left( k_n x \right) e^{\mp hx}.
\end{equation}
Physically, the non-vanishing value of $h$ indicates that the modes in the two non-Bloch bands are squeezed toward the two opposite ends of the lattice \SL{(skin effect)}. }
\SL{The effective dipole moment between OBC eigenstates $|u^R_{+,n}\rangle$ and $|u^R_{-,l}\rangle$ reads 
\begin{equation}
\mu^{(eff)}_{n,l} =   d_y \sinh(2 h)  \frac{(e^{h(1+N)}-(-e^{-h})^{(1+N)}) (\cos\theta_--\cos\theta_+)}{2(1+N)(\cosh 2h-\cos\theta_-)(\cosh 2h-\cos\theta_+)}  
 \end{equation}
with $\theta_\pm= k_n \pm k_l$. Note that $\mu^{(eff)}_{n,l}$ is \emph{non-vanishing} even for non-vertical \SL{transitions} $n\neq l$.
\SL{ This means that, under appropriate resonance forcing, ROs can be induced between non-vertical modes, as shown in Fig.3.}
\SL{Note that, in the large $N$ limit, $\Omega_R\sim e^{hN}/|n-l|^2$, which scales exponentially with $hN$ and to the inverse square of $|n-l|$. 
The exponential scaling with system size, arising from the skin effect,  provides the enhancement of the Rabi frequency, while the inverse power-law dependence on $|n-l|$ indicates that non-vertical transitions, with decreasing strengths as $|n-l|$ increases, are allowed when $h \neq 0$}. \SL{Examples of enhanced ROs in a lattice comprising $N=31$ sites for increasing values of the non-Hermitian parameter $\kappa_1 / \kappa_2$ are shown in Fig.4.}}

\begin{figure}[htbp]
  \includegraphics[width=140mm]{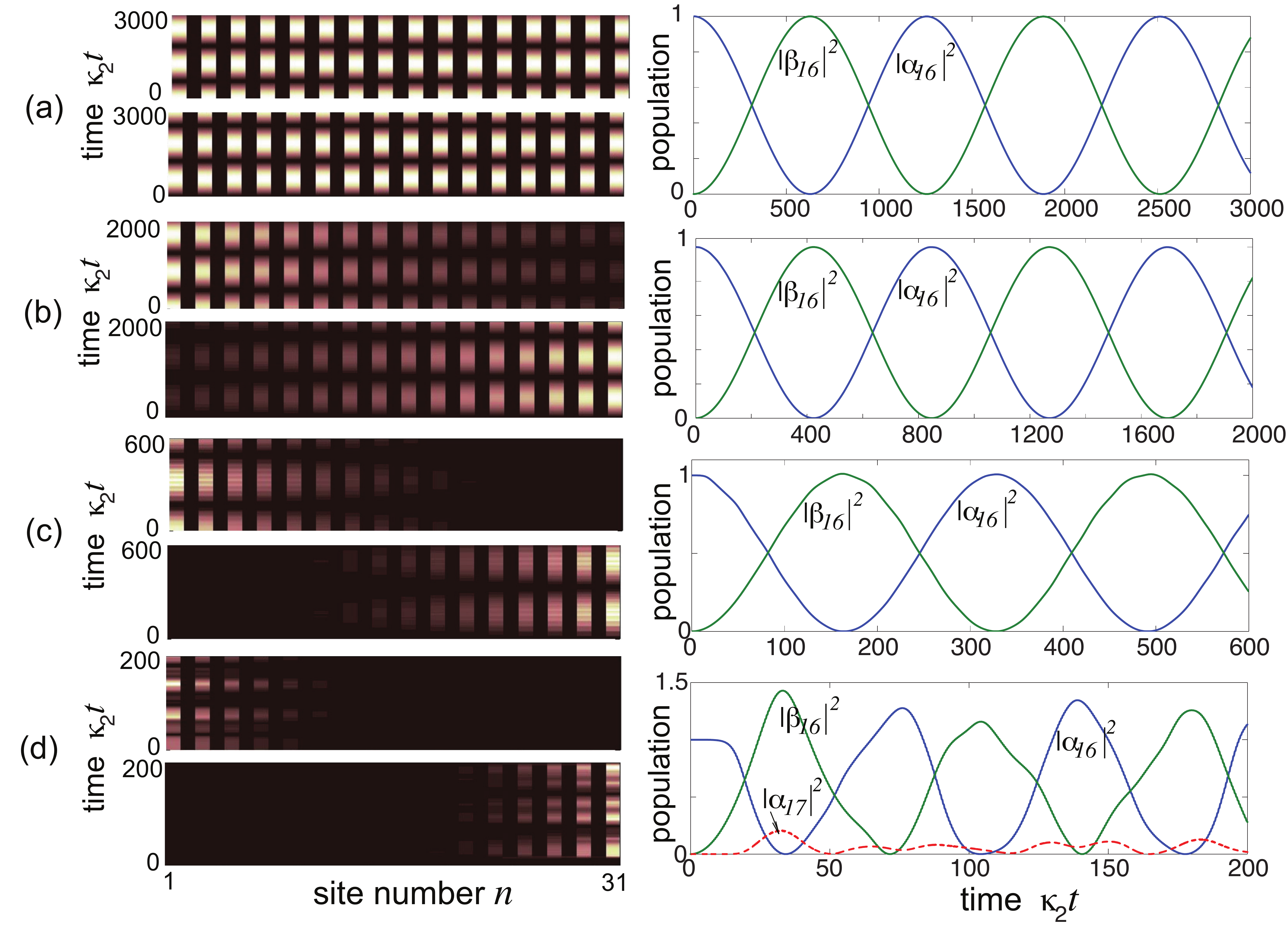}\\
   \caption{\SL{(color online) Enhanced RO frequency induced by skin effect and anharmonic ROs for increasing non-Hermitian parameter: (a) $\kappa_1 / \kappa_2=0$ (Hermitian limit); (b) $\kappa_1 / \kappa_2=0.05$; (c) $\kappa_1 / \kappa_2=0.1$; (d) $\kappa_1 / \kappa_2=0.16$. Other parameter values are: $N=31$ lattice unit cell, $\Delta / \kappa_2=3$, $\omega=2 \Delta$, and $F_0 / \kappa_2 =0.005$. 
   The initial state is the bulk eigenstate $| \mathbf{u}^{R}_{n,+} \rangle$ with quasi-momentum $k_n= n \pi /(N+1)$ and $n=(N+1)/2=16$.
   Left panels: temporal evolution of the real-space occupation probabilities $|A_n|^2$ and $|B_n|^2$ (upper and lower plots) on a pseudo color map. Right panels: evolution of the occupation probabilities $|\alpha_{16}|^2$ and $|\beta_{16}|^2$. In (d) the evolution of $|\alpha_{17}|^2$ is also depicted (as an example), indicating the emergence of mixed and anharmonic ROs.}}
\end{figure}
\begin{figure}[htbp]
  \includegraphics[width=100mm]{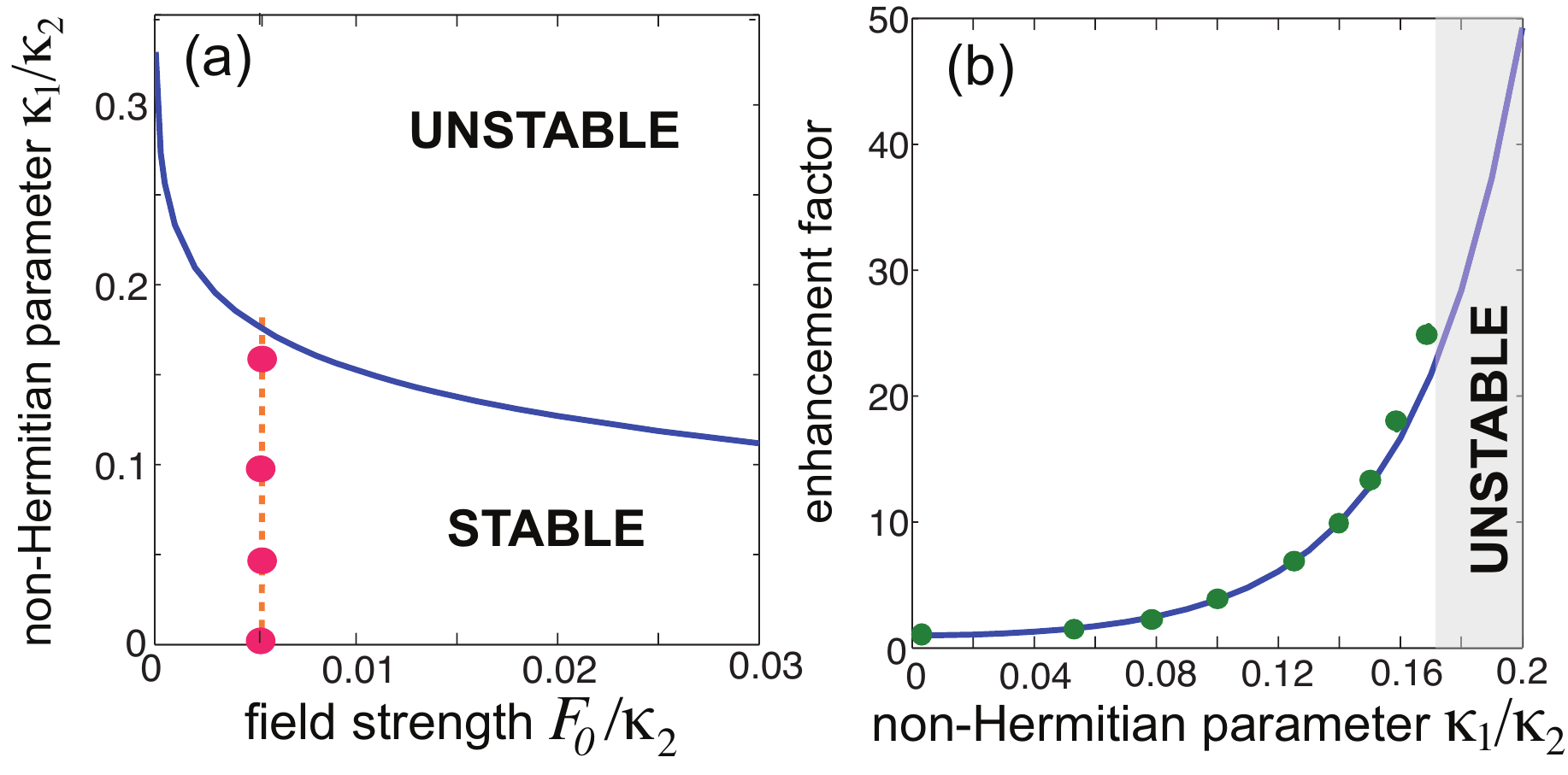}\\
   \caption{(color online) \SL{ (a) Stability domain in the $(F_0 / \kappa_2, \kappa_1 / \kappa_2$) plane for $\Delta / \kappa_2=3$, $\omega=2\Delta$, $N=31$. In the unstable domain  quasi energies are complex, corresponding to exponentially-diverging amplitudes. The circles  in the stable region correspond to parameter values used in the simulations of Fig.4. (b) Behavior of the effective dipole $\mu^{(eff)}_{n,n} / d_y$ versus the non-Hermitian parameter $\kappa_1 / \kappa_2$ for the same parameter values as in (a) and for the bulk mode corresponding to $k_n= \pi/2$, i.e. $n=(N+1)/2=16$. The solid curve shows the theoretical prediction [Eq.(11)], while the circles are obtained from the full numerical solution of Eqs.(5,6).}}
\end{figure}

\vspace{0.5cm}
\noindent{{\bf Stability of quasi-energies and anharmonic Rabi oscillations}\\
\\
Non-vertical transitions and boundary-induced dipole enhancement can invalidate the crude RWA, with the appearance of a complex quasi-energy spectrum, indicating that the system enters into an unstable regime and ROs are not anymore observed, as shown in the Methods. The onset of complex quasi energies, as determined by a Floquet analysis of Eqs.(5) and (6), can be regarded as a kind of parametric instability of the ac-driven system \cite{r35a,r35}}, as indicated by the appearance of unstable resonance tongues in the $(\omega,F_0)$ plane emanating from some of the transition frequencies $\omega_{nl}$ (see Figs. 7 and 8 in the Methods). The instability can be physically explained  from  the field-induced coupling between the two chains $H_+$ and $H_-$ that makes it possible a secular amplification of excitation along closed loops 
(Fig.6(c) in the Methods). We remark that breakdown of the RWA and the onset of unstable dynamics is a genuine {\em boundary-driven} non-Hermitian effect arising from non-vertical transitions, i.e. it is not related to counter-rotating terms like in the ultra strong coupling regime of light-matter interaction. An example of the stability domain in the $(F_0,h)$ plane, for fixed value of lattice site $N$ and modulation frequency $\omega$ far from any resonance tongue, is shown in Fig.5(a). To observe oscillatory (Rabi-like) dynamics, parameter should be chosen in the stable domain. The largest enhancement factor of effective dipole moment in RO is thus ultimately limited by the onset of the instability, as shown in Fig.5(b).  
Interestingly, as the system parameters are varied inside the stability domain to approach the stability boundary, ROs become highly anharmonic [see Fig.4(d)], a phenomenon which is clearly impossible to be observed in any Hermitian lattice.

\vspace{1cm}
\begin{large}
{\bf Discussion}\\
\end{large}
--------------------------------------------------------------------------------------------------------------------------------------------------------

 The fundamental processes of absorption and emission of quanta in crystals, as well as coherent Rabi flopping, are deeply modified by edge effects when considering non-Hermitian lattices displaying the skin effect. In particular, boundary-driven non-vertical transitions and anharmonic coherent Rabi flopping can be observed. Such results challenge the common wisdom that bulk coherent processes in crystals are largely independent of boundaries, indicating that skin effects not only question the bulk-boundary correspondence but also coherent bulk phenomena. Owing to the recent experimental progresses in the realization of synthetic non-Hermitian lattices displaying the non-Hermitian skin effect in photonic \cite{lei2019observation}, mechanical  \cite{ananya2019observation} and electrical circuit \cite{r31,tobias2019reciprocal} platforms, as well as pertinent theoretical advances in cold atom engineering \cite{li2019topology}, we expect that the disclosed distinctive physics of absorption and emission of  energy in non-Hermitian crystals could be experimentally accessible in the near future.

\newpage
\vspace{1cm}
\begin{large}
{\bf Methods}\\
\end{large}
--------------------------------------------------------------------------------------------------------------------------------------------------------
\vspace{0.5cm}
\noindent{{ \bf Generalized Brillouin zone and vertical transitions}} \\
\\
In the main text, we discussed vertical and non-vertical transitions between the effective chains governed by Hamiltonians $H_\pm$. Under PBCs, their eigenstates $u_{\pm,n}$ are labeled by the integer $n$ through the quantized lattice quasi-momentum $k_n=2\pi n/N$, and it is clear that vertical transitions refer to those connecting $u_{+,n}$ and $u_{-,n}$ of the same $n$ and thus quasi-momentum. Under OBCs, however, no well-defined lattice momentum exists, and below we shall discuss how the notions of vertical/non-vertical transitions can be extended more generally.\\
\\
{\it Generalized Brillouin zone for OBC systems}\\
\\
In the absence of the non-Hermitian skin effect, the OBC and PBC eigenstates can be mapped one another in a simple way, and it is physically expected that PBC transitions remain mostly unchanged upon the introduction of open boundaries in a large system. Indeed, for Hermitian lattices with NN hopping and vanishing asymmetry  ($h=0$), the OBC eigenstates [Eq.~(10)] are simply odd superpositions of PBC eigenstates with Bloch profiles $e^{\pm i k_n x}$, where $k_n=\frac{n\pi}{N+1}$, $n=1,...,N$ to give rise to $N$ unique OBC states (defined slightly differently from the PBC $k_n$). The introduction of the non-Hermitian skin effect ($h\neq 0$) amounts just to an exponential factor that can be obtained by deforming the quasi-momentum via $k_n\rightarrow h_n \pm ih$. In this simple example, it is clear that $n$ is still a well-defined label of the eigenstates, even though they are no longer in Bloch form.

More generally, there exists an almost 1-to-1 correspondence between the OBC and PBC eigenstates through the introduction of a generalized Brillouin zone (GBZ). Intuitively, most of the eigenvalues of the PBC spectrum will adiabatically ``flow'' en masse towards the OBC eigenvalues when the system is continuously interpolated between PBCs and OBCs i.e. by slowly switching off the end-to-end couplings, as detailed in Ref.~\cite{r24}. In other words, most of the PBC eigenstates, which will have evolved into OBC bulk eigenstates in the absence of the skin effect, will evolve together as boundary-localized skin eigenstates under the skin effect. The exceptions, if any, are topological eigenstates that evolve separately from the rest~\cite{r24,r26}. 

Recently, an ``unraveling'' picture for the GBZ was developed~\cite{lee2019unraveling} to explain how the most general set of skin eigenstates $u^\text{OBC}_n$ for the system with hopping asymmetry $h$ can be understood in terms of non-analytic complex momentum deformations of PBC eigenstates. This complex momentum lives in the so-called GBZ. First, we introduce the concept of the \emph{surrogate} Hamiltonian $\bar H$ which does not experience the skin effect, defined via a complex momentum deformation of the physical Hamiltonian $H$ in momentum space:
\begin{equation}
\bar H^\text{PBC}(k)=H^\text{PBC}(k-i\rho(k))
\label{eq1}
\end{equation}
\noindent Here $\rho(k)$ is the complex momentum deformation required such that there exist a double degeneracy in the decay lengths of the eigenstates i.e. $\rho(k)$ is determined by the condition that for each $k$, there exists another $k'$ such that the energy dispersion obeys $E(k+i\rho(k))=E(k'+i\rho(k'))$ with $\rho(k)=\rho(k')$~\cite{r24,lee2019unraveling}. In Hermitian cases, we always have $\rho(k)=0$ because as $k$ varies over a period, the energy dispersion $E(k)$ must always retrace itself before going back to its original value after a period in $k$.

To intuitively understand why the $\rho(k)=\rho(k')$ gives rise to a surrogate Hamiltonian that does not experience the skin effect, consider explicitly constructing wavefunctions that satisfy OBC from superpositions of the PBC momentum eigenstates. If the superposition coefficients were to converge in the thermodynamic limit, we need the presence of bulk-boundary correspondence i.e. the absence of the skin effect. At a particular energy $E$, OBC wavefunctions must satisfy two boundary conditions, namely that they vanish at both ends, i.e. at $x=0$ and at $x = N+1$. A superposition of at least two nonzero eigenstates is required for the wavefunction to vanish at $x=0$. For it to also vanish at $x=N+1$ in the thermodynamic limit of arbitrarily large $N$, another prerequisite is that both eigenstates must decay at the same rate, for otherwise one of them will be infinitesimally small compared to the other, and cannot possible cancel it off at $x=N+1$. This requirement that both eigenstates decay at the same rate is just that their imaginary momentum components are equal, i.e. $\rho(k_n)=\rho(k_m)$. This is thus also the condition for $\bar H$ to experience no skin effect.

Under OBC where momentum ceases to be a good quantum number, the complex momentum deformation $k\rightarrow k-i\rho(k)$ can be expressed as a similarity transform $S$ with a complex gauge field i.e. 
\begin{equation}
\bar H^\text{OBC}=S^{-1}H^\text{OBC}S
\label{eq2}
\end{equation}
An important corollary of this is that PBC Hamiltonians related by  Eq.~(\ref{eq1}) possess identical OBC spectra. Hence, to understand the OBC spectrum of a generic non-Hermitian Hamiltonian $H$, it suffices to determine that of its surrogate Hamiltonian $\bar H$, which is also almost equal to the PBC spectrum of $\bar H$ since the latter obeys the bulk boundary correspondence. In the thermodynamic limit where almost all states (except for isolated edge states) are skin states, the OBC spectrum $E^\text{OBC}$ is thus almost exactly indexed by $E^\text{PBC}(k_n-i\rho(k_n))$.

From Eq.~(\ref{eq2}), $u^\text{OBC}=S\bar u^\text{OBC}$, both with the same eigenenergy. Since $\bar u^\text{OBC}$ is not a skin state, it can be expanded in terms of the PBC eigenstates i.e. $\bar u^\text{OBC}_n=\sum_{n'}c_{nn'}\bar u^\text{PBC}(k_{n?})$ where $n,n'$ label the eigenstates and $c_{nn'}$ are coefficients that converge in the thermodynamic limit. Hence  
\begin{eqnarray}
u^\text{OBC}_n&=& S\,\bar u^\text{OBC}_n\notag\\
&=& S\sum_{n'}c_{nn'}\bar u^\text{PBC}(k_{n^{\prime}})\notag\\
&=&\sum_{n'}c_{nn'}\bar u^\text{PBC}(k_{n^{\prime}}+i\rho(k_{n^{\prime}})).
\label{eq3}
\end{eqnarray}
For our case of 1-band Hamiltonians $H_\pm$, $k_n=\frac{n\pi}{N+1}$, $\bar u^\text{PBC}(k_{n^{\prime}})\propto e^{ik_{n^{\prime}}x}$ and $\rho(k_{n^{\prime}})=h\,\text{sgn}(\sin k_{n'})$. Together, these yield Eq.~(10), with biorthogonal normalization imposed.\\ 
\\
{\it Ansatz for well-defined vertical transitions}\\
\\
In the main text, we have used the Ansatz $H_\pm(k)=Q_\pm H_H(k-ih_\pm)$ for producing $H_\pm$ with real and identical OBC spectra. Since the $ih_\pm$ complex deformation can be implemented as a similarity transform $S=\text{diag}(1,e^{h_\pm},e^{2h_\pm},...,e^{(N-1)h_\pm})$ in real space under OBCs, it is evident that $H_\pm(k)$ must possess identical OBC spectra up to a sign $Q_\pm$ [see also Eq.~(\ref{eq2})]. 

More generally, constant complex momentum deformations $k\rightarrow k+ik_0$ ($k_0$ a constant) generate an equivalence class of models with identical OBC spectra. In particular, since $H_H(k+i0)$ is Hermitian, the OBC spectrum must be real. This construction holds for multiband models as well, where we can further generalize the definition of $H_\pm$ to $H_\pm(k)=Q_\pm U^{-1} H_H(k-ih_\pm)U$ where $U$ is a unitary transformation. 

The eigenstates of $H_\pm^\text{OBC}$ can be inferred from Eq.~(\ref{eq3}) with $\rho(k)=-h_\pm$, where $\bar H$ is just $H_H$. Importantly, it is clear how the $n$-th eigenstates of $H_\pm$ correspond with each other. Again, specializing to our 1-band model, we shall recover the expression of Eq.~(10).

Having established the correspondence between the spectra and eigenstates of $H_\pm$, the notion of vertical ($n=m$) vs. non-vertical ($n\neq m$) transitions is well-defined.\\
\\
{\bf Quasi-energy spectrum and stability of Rabi oscillations}\\
\\
 \begin{figure}[htbp]
  \includegraphics[width=168mm]{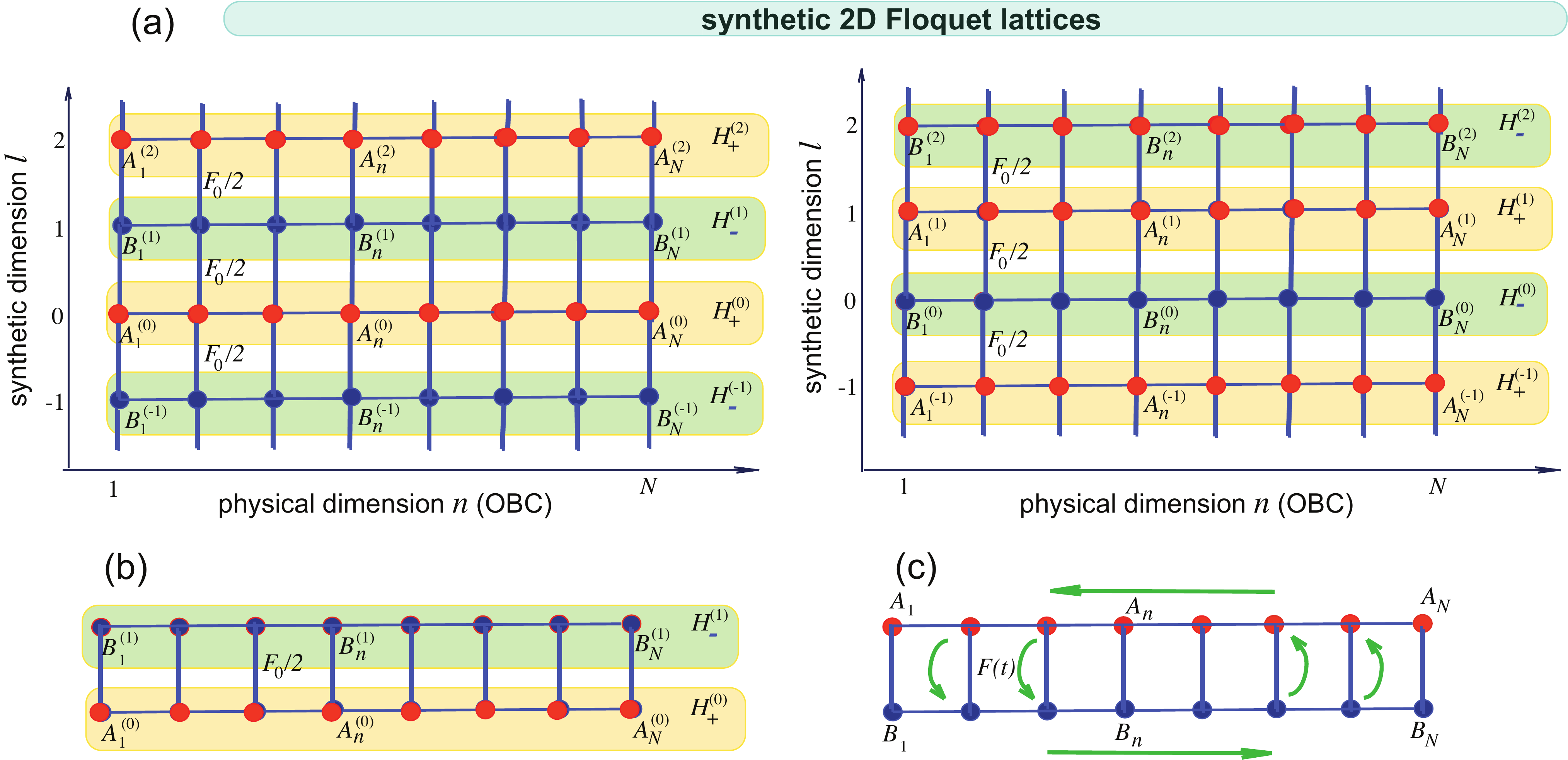}\\
   \caption{(color online) \SL{ (a) Equivalent decoupled 2D Floquet lattices of the ac-driven system. (b) Reduced model under wide gap and near-resonant forcing. (c) Simple physical explanation of the onset of instability. In the absence of forcing, excitations are squeezed toward the two opposite edges of the upper and lower chain, and amplification is prevented by the OBC. As the ac field is switched on, chain coupling realizes closed loops (schematically depicted by arrows), along which excitation can be secularly amplified, resulting in a complex quasi energy spectrum.}} \label{FigSupp1}
\end{figure}
The quasi-energy spectrum of the ac-driven lattice under OBC can be determined by a standard Floquet analysis of Eq.(2), or likewise of Eqs.(5) and (6). Let as assume a sinusoidal driving field $F(t)=F_0 \cos (\omega t)$, and let us search for a solution to Eq.(2) in the form of a Floquet eigenstate, i.e. 
\begin{equation}
\mathbf{A}(t)= \sum_{l} \mathbf{A}^{(l)} \exp(-i \mu t+i \omega l t) \; , \; \;  \mathbf{B}(t)= \sum_{l} \mathbf{B}^{(l)} \exp(-i \mu t+i \omega l t) \label{Eqpara1}
\end{equation} 
where $\mu$ is the quasi-energy, which is uniquely defined in the range $(-\omega/2, \omega/2)$.  Substitution of the Ansatz (\ref{Eqpara1}) into Eq.(2) given in the main text yields the 
Floquet chains
\begin{equation}
\mu \mathbf{A}^{(l)}  =   H_+^{(l)} \mathbf{A}^{(l)}+\frac{F_0}{2} \left( \mathbf{B}^{(l+1}+ \mathbf{B}^{(l-1) } \right) \; , \; \; \;
\mu \mathbf{B}^{(l)}  =  H_-^{(l)} \mathbf{B}^{(l)}+\frac{F_0}{2} \left( \mathbf{A}^{(l+1}+ \mathbf{A}^{(l-1) }\right) 
\end{equation}
 \begin{figure}[htbp]
  \includegraphics[width=140mm]{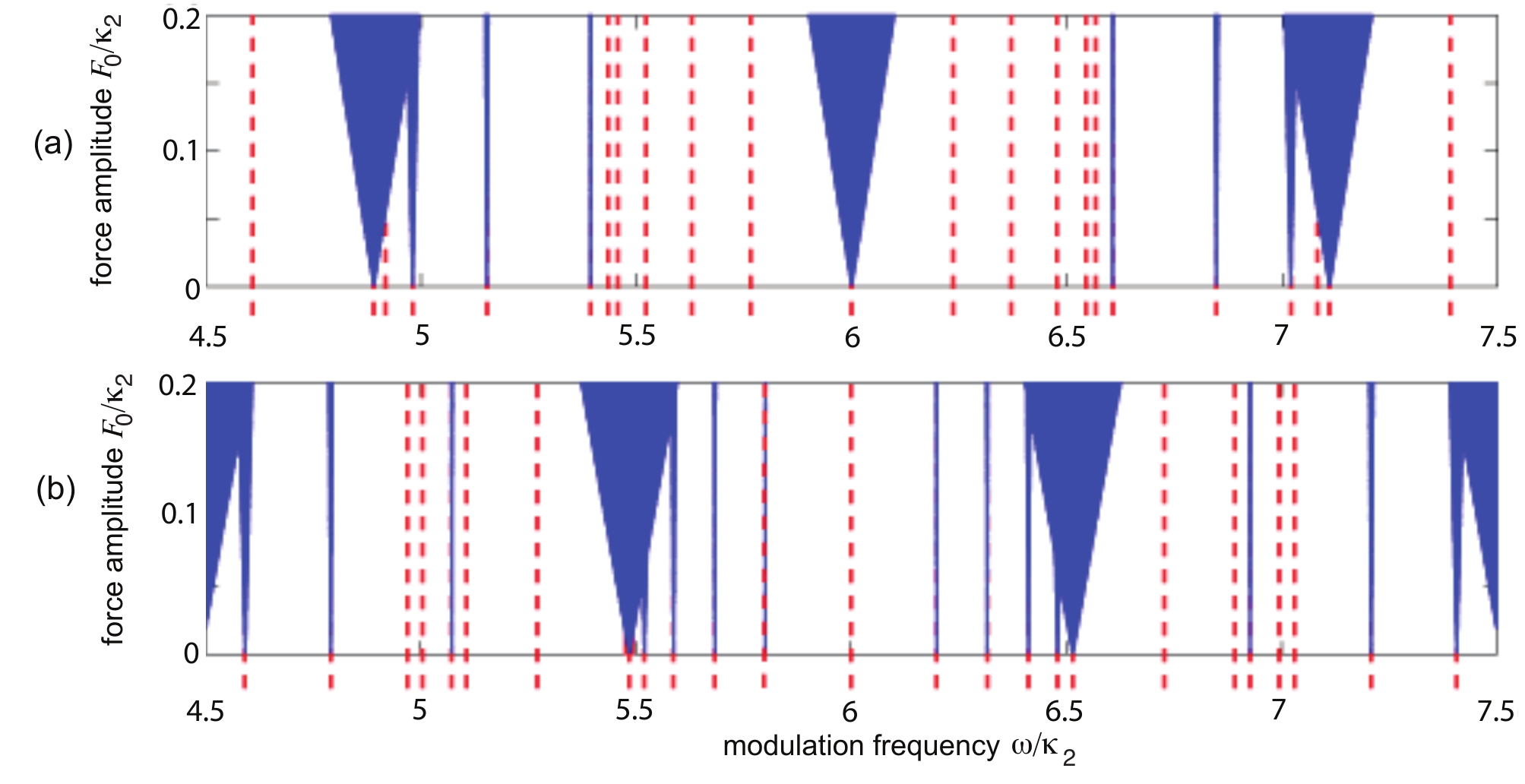}\\
   \caption{(color online) \SL{ Resonance tongues (shaded areas), showing the onset of complex quasi-energies, in the $(F_0, \omega)$ plane for (a) $N=10$, and (b) $N=11$ sites in the lattice. Other parameter values are $\kappa_1 / \kappa_2=0.1$ and $\Delta / \kappa_2=3$. The dashed vertical lines in the plots show the position of the various resonance transitions $\omega_{n,l}\equiv (E_{+,n}-E_{-,l})$ between the OBC energy levels in upper and lower bands. Note that the unstable resonance tongues emerge from some (but not all) the resonance transitions.} } \label{FigSupp2}
\end{figure}

 \begin{figure}[htbp]
  \includegraphics[width=150mm]{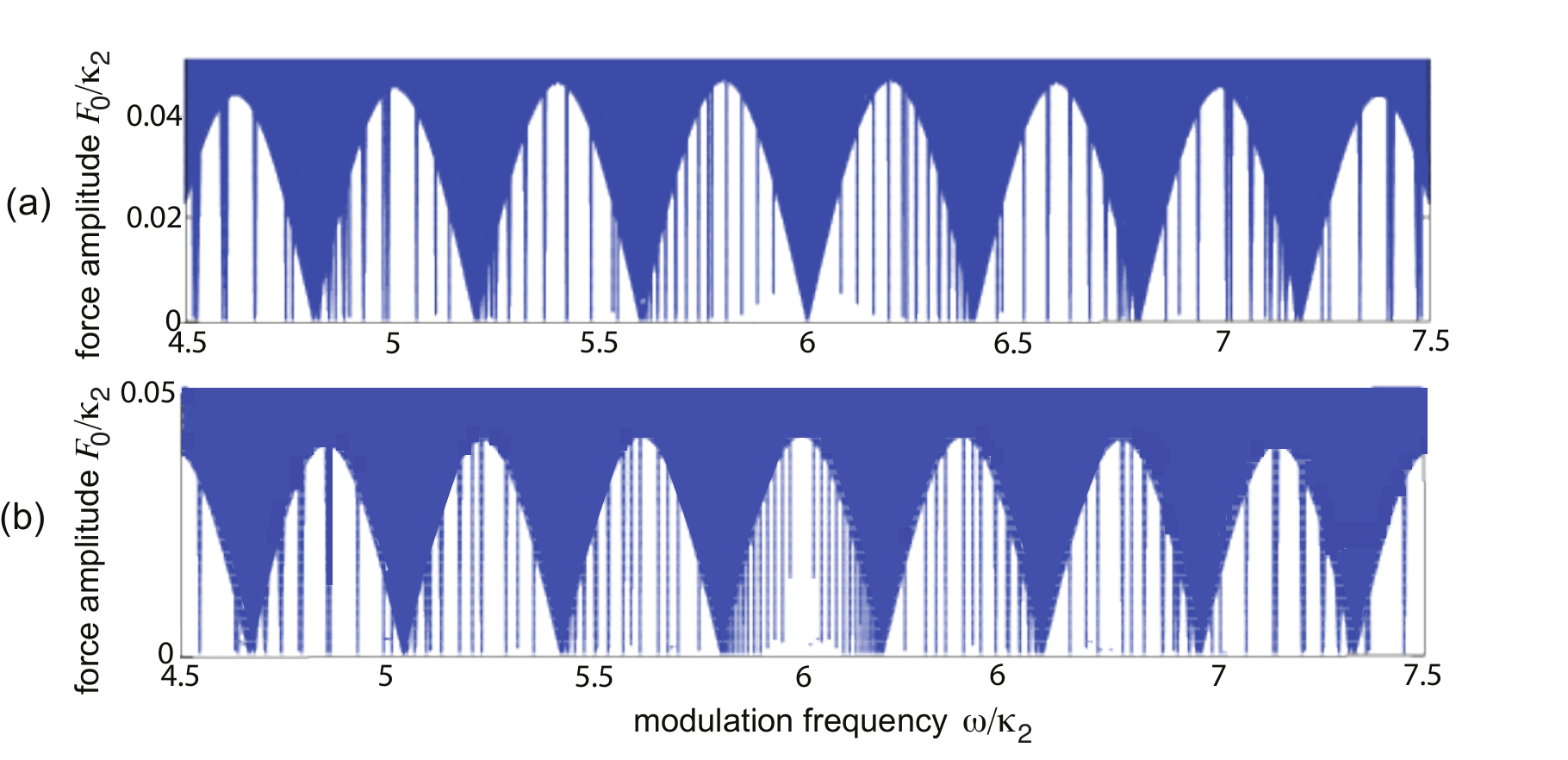}\\
   \caption{(color online) \SL{ Same as Fig.S2, but for (a) $N=30$, and (b) $N=31$ sites in the lattice. For the sake of clearness, the vertical dashed curves, corresponding to the resonance transitions, are not displayed.}} \label{FigSupp3}
\end{figure}
\vspace{1cm}
where we have set $H_{\pm}^{(l)} \equiv H_{\pm}+l \omega$. The quasi energies $\mu$ can be thus viewed as the eigen-energies of two decoupled static 2D lattices (stripes), in which one dimension (labeled by the index $l$, varying from $-\infty$ to $\infty$) is a synthetic (frequency) dimension while the other dimension (labeled by the index $n$, varying from $n=1$ to $n=N$) is the physical space, as schematically shown in Fig.\ref{FigSupp1}(a). A special case is the one in which the two non-Bloch bands, defined by the Hamiltonians $H_{+}$ and $H_{-}$ under OBC, are entirely real and spaced by a wide gap $2 \Delta$, like in the model of Fig.1. Under near-resonant driving $ \omega \simeq 2 \Delta$ and assuming weak forcing, application of standard multiple-scale asymptotic analysis (see e.g. \cite{FloquetLonghi}) indicates that the full Floquet chains of Fig.\ref{FigSupp1}(a) basically decouple into equivalent quasi 1D static lattices, as shown in Fig.\ref{FigSupp1}(b). Thus, at leading order the $2N$  quasi-energies $\mu$ of the ac-driven system are obtained as the eigenvalues of the linear system 
\begin{equation}
\mu \mathbf{A}^{(0)}=H_+ \mathbf{A}^{(0)}+ \frac{F_0}{2} \mathbf{B}^{(1)} \; ,\;\;\; \mu \mathbf{B}^{(1)}=(H_-+\omega) \mathbf{B}^{(1)}+ \frac{F_0}{2} \mathbf{A}^{(0)}.
\end{equation}
Let us specialize the general theory to the model discussed in the main text [Fig.1 and Eq.(8)]. For given values of the band spacing $2 \Delta$ and non-Hermitian deformation parameter $h$, the quasi energy spectrum $\mu$ turns out to be strongly sensitive to the modulation frequency $\omega$, force strength $F_0$, and site number $N$ (in particular $N$ odd or even). Extended numerical simulations, based either on diagonalization of the full Floquet chains [Fig.\ref{FigSupp1}(a)]  or the reduced quasi 1D lattice [Fig.\ref{FigSupp1}(b)], indicate the existence of multi-branch instability domains in $(\omega,F_0)$ space made of wide and narrow resonance tongues, where the quasi energy spectrum becomes complex. In such regions the amplitudes exponentially grow in time and the observation of oscillatory (Rabi-like) dynamics is thus prevented. Figures \ref{FigSupp2} and \ref{FigSupp3} show typical behaviors of resonance tongues for  $N$ odd and $N$ even. Interestingly, the unstable resonance tongues emerge from  some (but not all) of the transition frequencies $\omega_{n,l}\equiv (E_{+,n}-E_{-,l})$ between OBC energy levels in upper and lower bands. Physically, the onset of the instability can be readily explained from the diagram of Fig.\ref{FigSupp1}(c). Forward (backward) propagating excitations in the upper chain are attenuated (amplified) by the imaginary gauge field $h$. The opposite holds for the lower chain. In the undriven case ($F=0$), the two chains are decoupled. Owing to the OBC, under steady state the excitations in each chain are squeezed at opposite edges and can not grow/decay anymore: this means that the energy spectrum remains entirely real despite the non-Hermitian term $\kappa_1$ in the Hamiltonian. When the ac force is switched on, closed loops between upper and lower chains, with an overall net amplification of excitation per round trip, is allowed by the vertical solid bonds in Fig.\ref{FigSupp1}(c), resulting in complex quasi-energies. In other words, the ac force effectively changes boundary conditions allowing periodic circulation of excitations between the two chains.


\begin{thebibliography}{31}
\bibitem{note1}
We have $\Gamma_{n,l}G_{l,n}=\langle u^L_{+,n}|u^R_{-,l}\rangle\langle u^L_{-,l}|u^R_{+,n}\rangle=\text{Tr}[P^+_nP^-_lP^+_n]=\text{Tr}[P^+_nP^-_l]$ since $P^+_n$ is idempotent.
\end{thebibliography}
\end{document}